\documentstyle[12pt]{article}
 \newcommand \be {\begin{equation}}
\newcommand \bea {\begin{eqnarray} \nonumber }
\newcommand \ee {\end{equation}}
\newcommand \eea {\end{eqnarray}}
 
 \newcommand \s {\sigma}

 \newcommand \g {\gamma}

 \newcommand \al {\alpha}
 \newcommand \N {{\cal N}}

\newcommand \lan {\langle}
 \newcommand \ran {\rangle}
\newcommand{\bi}{\bibitem}
\newcommand{\bit}{\begin{itemize}}
\newcommand{\eit}{\end{itemize}}

\begin{document}

\title{Slow dynamics in glasses} \author{  Giorgio Parisi
\\
Dipartimento di Fisica,
Universit\`a {\sl La  Sapienza}\\ INFN Sezione di Roma I \\ Piazzale
Aldo Moro, Roma 00187}
\maketitle

\begin{abstract} We will review some of the theoretical progresses that
have been recently done in the study
of slow  dynamics of glassy  systems: the general techniques used
for
studying the dynamics in the mean field approximation and the
emergence of a
pure dynamical transition in some of these systems. We show how the
results
obtained for a random Hamiltonian may be also applied to a given
Hamiltonian. These two results open the way to a better
understanding of the glassy transition in real systems.
\end{abstract} 

\section {Introduction}
Many systems (among them glasses, rubber, spin glasses...) show at low
temperature a very slow  approach to equilibrium. Our aim is to
understand
this kind of behaviour, in  particular the peculiar properties of the
glass transition.

At the present moment in the framework of the mean field
approximation the static of spin glasses is well understood. We start
to have a reasonable understanding of the dynamics. We hope that
these progresses will lead to a better understanding of the behaviour of
glasses. However this goal seemed to be unreachable for two reasons.
\bit
\item  (a) The behaviour of  spin glasses and real glasses is rather
different near the transition.

 In real glasses a very interesting  phenomenon
happens: if  the temperature is  decreased fast enough, the system
goes into an  amorphous state that  has an extremely large mean  life;
moreover
 under very slow cooling  some of these systems go into an ordered
crystal phase. The  crystal
phase does not exist in general, it exists  only if the parameters of the
interactions among the atoms  are
{\sl intentionally} chosen in such a way that this phase is energetically
favoured. Independently from the
possible  existence of the crystal phase, the dynamics in the glassy phase
becomes extremely slow at
low  temperature.

Moreover the internally energy in the amorphous phase depends on the
cooling speed in a strong way, also for very slow cooling. These
effects have never observed in spin glasses. The spin glass transition can
also be characterised by the divergence of a static quantity (the non
linear
susceptibility) while no anomalous behaviour has been observed in the
static quantities in real glasses.

\item (b)
{}From the theoretical point of view the two systems seem to be
extremely
different.

Spin glasses are systems with a random Hamiltonian. They are studied
by
averaging over the disorder in the Hamiltonian  using {\sl ad hoc}
methods, e.g. the replica technique. Real glasses have a given
Hamiltonian
and the disorder is not present in the Hamiltonian, but comes out from
the
freezing of some of the relevant degrees of  freedom.
 \eit

Recently there have been some progresses which show that the previous
difficulties are less serious than it was believed.
\bit
\item (a) It has been found that in some disordered systems (e.g.
modified
spin glasses) there is a dynamical transition which is characterised by
the
divergence of the time needed for equilibrate the system. Near this
transition temperature the static quantities do  show no anomaly.
The
temperature at which the equilibration  time diverges is higher than the
temperature at which a transition is present for the static
quantities. This static transition cannot be observed because the system
takes to much time to equilibrate.

\item (b) In some cases the replica method
can be applied also to system without intrinsic disorder in the
Hamiltonian.
The behaviour of those system is very similar to that of really
disordered
systems, apart from the possible existence of a low energy ordered
crystal
phase
\eit

 Some of these effects cannot be understood  using only the tools of
standard equilibrium statistical  mechanics, because they are non-
equilibrium
phenomena. In this note I  will present a coherent
picture  of the dynamics of these systems.  The results  have been
obtained in these recent years
mainly in the  framework of spin glasses theory  \cite{MPV,PB2} and I
will show how they can be extended to other
systems like glasses.  Some of the statements are well proven, while
other are still conjectural.

In the second section of this note I will give a general
qualitative description of the  dynamics based on the
hypothesis that systems evolve in time jumping from one to an other
quasi-equilibrium state. In the
third  section I will review some very interesting results obtained by a
direct computation of the
dynamical  evolution of the system using more powerful methods; I will
also compare these results
with those  obtained in the previous section. Finally in the fourth section
I
will address to the old
problem of  comparing the behaviour of a system with random
Hamiltonian (like spin glasses), with
that of system  with a fixed Hamiltonian (like glasses). This
comparison will be done in a
model system: one finds  that the properties of the system with fixed
Hamiltonian are very similar
 to that of
the systems with random Hamiltonian. The only difference is the
 possible existence of  a crystal phase for specific choices of the
parameters of the fixed Hamiltonian; the crystal phase does not exists
for
the random Hamiltonian. Some brief conclusions are presented at the
end.

\section{Local Equilibrium States}

We  consider Ising spin models in which there  are $N$
variables  $\s_i$, which take
the  values $\pm 1$ \footnote{Similar consideration can also be done for
interfaces or manifolds in a
random medium \cite{FM}.} There are many possible kinds of
Hamiltonian that we can write down. In
the simplest case  the interaction involves only two spins,

\be
H_J(\s)=- \sum_{i,k} J_{i,k} \s_i\s_k,\label{SPIN}
 \ee
and all the pairs $i,k$ are equivalent.

 If the $J$ are randomly distributed Gaussian variables, with  variance
$1/N$, we obtain the SK model,
otherwise one has a different model.

At low temperature  these models have a corrugated free energy
landscape, with many local minima,
separated by high barriers. In this situation  general
arguments imply  a very slow dynamics
because the system may be trapped in a valley and it  takes quite a
long time to escape from it.

The  jumping from a valley to an another valley is a controlled by the
height of the barriers and
the time  needed (neglecting prefactors) is

\be \tau=\exp(\beta \Delta F), \ee
where $ \Delta F $ is
the minimum barrier in free energy that the system has to cross in
going from  one valley to an
other valley \footnote{This is not the only method for  having a slow
dynamics. For  alternative
possibilities see \cite{CKP,VI}.}.

The best characterisation of a valley
 \footnote{Sometimes one uses the
terminology {\sl local equilibrium state} or {\sl quasi-equilibrium
state}.}
is a  region of the phase space in  which the system spends a long time.
In this case it reasonable to  define the local
magnetizations in the  valley $\al$ as \be m_i^\al = \lan \s_i \ran _\al, \ee
 where the average is taken inside the valley $\al$.

In long range models these magnetizations satisfies the mean field
equations. Neglecting the Onsager reaction field
\footnote{If we add the Onsager reaction field
we obtain the TAP equations.}, they are
 are
 \be m_i= th (\beta \sum_k J_{i,k} m_k) \ee
Alternatively we can define a free
energy as function of the $m_i$: \be F[m]=  - \sum_{i,k} J_{i,k} m_i
m_k  -T \sum_i s(m_i), \ee
where the local entropy is simply given by
 \be s(m) = -{1+m \over 2}
\ln({1+m \over 2})-{1-m \over
2} \ln({1-m \over 2}). \ee
The local minima of the free energy  $F[m]$ are
 solutions of the mean field equations .

The free energy landscape can be characterised by the structure of the
set of the solutions (which
we  will label by Greek indices). Generally speaking the most important
parameters are the free
energy of  the solution ($f^\al$, the local magnetizations $m^\al_i$ in a
given solution, the
overlap among two  solutions $\al$ and $\g$ ($q_{\al,\g}={1\over N}
\sum_{i=1,N} m^\al_i m^\g_i$) and the self  overlap ($q_{EA}=
q_{\al,\al}$), which in most of  the models is independent from the
solution.

At equilibrium it is reasonable to assume that different valleys may be
populated, and the
probability  that the system is in one of this valley  is given by
\be
w_\al \propto w(f_\al)
\equiv \exp (-\beta f_\al).
 \ee
 It is evident that
 \be
\sum_\al w_\al=1. \label{NORM}
 \ee
 In many
disordered systems the number of valleys as function of the free energy
($\N(f)$) increases  as
\be \N(f) \sim \exp ( y (f-f_0)+O(({f-f_0)^2\over V}), \ee
in the  region where
\be
1<<f-f_0<<V
\ee
where $V$ is the volume of the system.

At a given temperature there are two possibilities
\begin {itemize}
\item{(a)} If $y<\beta$ the
integral
 \be \int df \N(f) w(f) \ee
is dominated by $f$ near $f_0$. Only
few valleys dominate the sum in  equation(\ref{NORM}), although an
infinite  number of them give a non zero
contribution. In this  situation we say that the replica symmetry is
broken.

\item (b) If $y>\beta$ the integral
 \be \int df \N(f) w(f) \ee
would be divergent if we neglect terms of $O(({f-f_0)^2\over V})$. In
reality the integral  is  dominated by the region where $f-f_0$ of order of
the size $V$ of the system. In this case the number of valleys which
dominates the sum in equation
(\ref{NORM}) is exponentially large and each of the valley has a weight
which is exponentially
small. In this case the replica symmetry is not broken.
\end{itemize}

Let us consider the case where  the valleys are separated by very high
barriers (e.g. diverging with
$N$) in the region where the temperature $T$ is smaller than $T_D$.
Depending on the nature of the
problem we may enter or in the region (a) or (b) when we decrease the
temperature from above to below $T_D$.

If we enter in the  region (a), as it
happens in the  usual SK model for spin glasses, a phase transition is
present from  the equilibrium point of
view at  $T_D$.

 On the contrary, as happens in other spin glass models
\cite{Cri}-\cite{CuKu2}, if we enter in the region (b), no phase
transition is present from the  equilibrium point of view at $T_D$, and a
static transition is present only at smaller  temperature, $T_R$, where $y$
becomes smaller that $\beta$  and we pass from region (a) to
region  (b). This transition can be easily understood. At temperatures
greater that $T_R$ the valleys populated at equilibrium have a free
energy
density greater than minimal possible because of entropic effects. The
relevance of these entropic effects disappear by decreasing the
temperature and for temperatures less than $T_R$ the valleys have the
smallest possible free energy.

 In this last case we have two transition one for the statics and
the other for the dynamics. If we quench the system at temperature
smaller than $T_D$ coming from an high temperature region, the
internal energy (for an infinite system) does not go to the equilibrium
value and the system remains trapped in a metastable state
\cite{Cuku,Cuku2}.

In this case for the infinite system independently from the speed of cooling
we always find the energy of the metastable states and therefore
we
do not observe any strong dependence of the energy on the cooling.
This is an
artefact of the mean field approximation, which is correct only for
infinite
range forces. We shall see later how this behaviour may be changed for
more
realistic model.

It is also possible that there is an other isolated solution to the mean
field equation, with a free
energy   density $F_C$ smaller than $F_0$ and there are no solutions in
the region
\be
 F_C<F<F_0.
\ee
In many cases this isolated solution describes an highly ordered
state,
which  we call the crystalline
state. The  existence of this states does not change the properties of the
system in the region
where $f>f_0$.

{}From this point of view in order to compute the approach to
equilibrium  one should evaluate the
free  energy barriers which separate one valley from an other valley.
This computation is rather
difficult,  especially if we take care that the system is still slightly out of
equilibrium. However
in the next section  we shall see that a direct computation of the non
equilibrium properties can be
done.

\section {The non equilibrium equations}

It was found quite recently \cite{Cuku,Cuku2} that the non equilibrium
behaviour of the system can be described
directly  for the infinite system (i.e. after having taken the limit $N \to
\infty$), by introducing
the average correlation  function and the response function defined as
\bea
C(t,t')=\lim_{N\to\infty}{\sum_{i=1,N} \s_i(t) \s_i(t') \over N},\\
G(t,t')=\lim_{N\to\infty}{\sum_{i=1,N} {\delta \s_i(t') \over \delta
h_i(t)} \over N}.
\eea
In the
equilibrium regime time translation invariance implies that these
 functions depend only
on the time difference. We consider here the case where the system at
time zero starts from a
random  configuration. Only positive times are possible and time translation
is explicitely broken.

 Closed equation can be written for these two  functions.  They are
\bea {\partial
C(t,t')\over \partial t'}= E_C[C,G],\\
 {\partial G(t,t')\over \partial t'}= E_G[C,G], \eea
 where
$E_C[C,G]$ and $E_G[C,G]$ have an explicit form (non local in time)
which depend on the  problem. In
some case one can expand $E_C[C,G]$ and $E_G[C,G]$ in powers of
$C$ and $G$.  The  solution of these
equations can be computed numerically and one can obtain a great
amount
of  information in this way.

 From the analytic point of view one can study these equation in the
adiabatic approximation, where
one set to zero the time derivative. This approximation is justified in the
large time region. The
resulting equations are non trivial; they are
\be
 E_C[C,G] = E_G[C,G]=0.
\ee
The solution of
these equations can be simplified \cite{FM,Cuku,Cuku2} by noting that
they are {\sl reparametrization  invariant}, i.e.
if $C$ and $G$ are a solution also the functions
\bea
 C_h(t,t')=C(h(t),h(t')), \\
G_h(t,t')=G(h(t),h(t')){dh(t')\over dt'}, \eea
are an other solution of the adiabatic equations, for
an arbitrary choice of the function $h$.

Reparametrization invariance strongly simplifies the study of the
adiabatic equations and many
results  can be obtained in this limit. In some case it can be proved that
for large times the
internal energy tends to the  equilibrium value, while in other case finds
that there
is dynamical transition at a temperature $T_D$.  At lower temperature
one finds that  the dynamical
energy does {\sl not} tend at large times at the  equilibrium value and
therefore metastable states
are present.

The evaluation of the reparametrization invariant quantities morally
corresponds to the evaluation
of  the properties of the solutions of the mean fields equations of the
previous section, although
it contains  more information.

The more difficult part, which at the present moment we can do only
numerically, consists in
computing quantities that are not reparametrization invariant, as the
time dependence of the energy.
This computation morally corresponds to the evaluation of the barriers
separating the solutions of
the  mean fields equations of the previous section and it is not a surprise
that it turns out to be
much more  difficult. Technically one ends up with a well defined and
difficult mathematical
problem, very similar  in spirit, but more complicated, of the non linear
velocity selection
problem, which has been widely  studied in the past.

It is extremely satisfactory that the very difficult problem of computing
analytically the non
equilibrium dynamics in these systems is now under control and I am
convinced that the mathematical
difficulties may be surmounted, may be with some help from our more
mathematically minded  friends.

The results obtained from this dynamical approach have the advantage to
be easily compared with those
obtained experimentally in spin glasses, where the condition $N$ very
large with respect to $t$ is
certainly satisfied. A very interesting phenomenon which appears is
aging,  i.e. the dependence of
the   experimental results on the age of the system
\cite{Bo1,Bo2,Cukuri}. A detailed discussion of this point would make
this note  too long.

\section{Glasses in the mean field approximation}
In spin glasses the
Hamiltonian is random as an  effect of quenched random
disorder. In real glasses the  Hamiltonian is not random and the
quenched disorder is dynamically
generated at low temperature. We  can  ask how much of the
qualitative and quantitative
results which have obtained in spin  glasses may be transferred to
glasses.

In order to understand this point we have started to study models in
which the Hamiltonian does not
contain quenched disorder and to compare the results with those
coming
of random Hamiltonian \cite{MPR}.

Our strategy is the following. We want to study the properties
of a given Hamiltonian
$H_G$ which is not random. We consider a class of Hamiltonians
$H_R$, of which $H_G$ is a
particular case. We choose the class $H_R$ in such a way that the
statistical properties of $H_G$
and  that  of a  generic Hamiltonian in $H_R$ are as similar as possible.
In the best case we can
obtain that the two  corresponding free energies coincide in the high
temperature expansion \footnote{In
general the behaviour of  the system can be better controlled in the high
temperature phase.}.

After having constructed $H_R$ in an appropriate way, we can suppose
that  the statistical
properties   of $H_G$ and $H_R$ are the same or, if they are different,
we can construct a
perturbative expansion  which compute this difference. It is clear that
this approach may be
successfully in the high  temperature region (more or less by
construction) and it may also
reproduce the behaviour in the  glassy region, included the dynamic and
static transitions. However
it is cannot certainly reproduce the  possible existence of a crystal phase.

I will present now a simple model in which this approach works very
well at all the temperature and
it  misses only the crystal phase, which exists only for {\sl intentionally}
chosen Hamiltonians
$H_G$ \cite{MPR2}. The  Hamiltonian is is the same  as in
eq. (\ref{SPIN}).

In the case of $H_G$ we have
 \be
 J_{i,k}= N^{-1/2}\sin({2 \pi i k \over N}),  \ee
while in the case of
$H_R$ we have that $J$ is a random orthogonal symmetric matrix, i.e.
a  random symmetric matrix which satisfies the constraint
\be
 \sum_k J_{i,k}J_{k,j} = \delta_{i,j}. \ee
It is easy to check that the first Hamiltonian is a
particular realization of the second one.

A details computation (partially analytic and partially numeric) for the
Hamiltonian $H_G$ shows
that  in the low temperature phase there are two different limits when
$N$ goes to infinity,
one  for generic  $N$ and an other for $N$ odd, such that $p=2N+1$ is
prime. Only in this second case
there is a crystalline phase  at low temperature. At all
temperatures the model for generic
$N$   behaves in the
same way as the model described by  $H_R$ and  it undergoes   a replica
symmetry breaking transition. On the contrary the dynamical
glassy transition is present for all $N$.
 The strategy of computing the
properties of a given system by  approximating it by a random system  works
very well in the glassy region and it misses the crystal  phase, which
exists only for non generic  values of $N$.

There are many other system which can be studied using this stategy
\cite{MIGLIO,ANA}, but I will not discuss them.

\section{Conclusions}
 We have seen that we begin to control the off
equilibrium dynamics in many
model systems. There are  cases in which we have a glass transition with
metastability. This
feature is present only in the mean  approximation, which is correct in
the infinite range limit.

A crucial and open problem is to understand how this feature of the
infinite range approximation
appears in more realistic finite range models, in which metastable states
cannot exists. Let us describe a possible scenario for the (fragile) glass
transition.

 The correlation time $\tau$ diverges algebraically at $T_D$, which is
not the
situation for real glasses where a divergence of the form
\be
\tau \propto \exp({A \over (T-T_G)})
 \ee
is observed.

On the other hands metastability is present in the mean field
approximation
as soon as $T<T_D$. Generally speaking metastable states do decay in
short
range model by tunnelling effects. If we apply these ideas to the present
case we find that
\be
\tau \propto \exp (g(T)),
\ee
where $g(T)$ is a smooth function, which quite likely diverges at the
temperature $T_R$ as
\be
g(T) \propto (T-T_R)^{-\g}.
\ee
{}From this point of view the glass transition temperature should be
identified
with the temperature at which in the mean field approach the replica
symmetry is broken. The dynamical transition found in the mean field
approach is no more a real transition and denotes the onset of very slow
dynamics.

   This line of thinking should be
seriously investigated and one should develop the  theoretical
techniques needed to compute the function $g(T)$. The dependence of
the exponent $\g$ on the dimensions of the space is a rather interesting
issue, which unfortunately it has not been studied up to now. Experiment on
four dimensional glasses are notoriously rather difficult to be done,
however  the comparison of numerical simulations for three and four
dimensional  glass models should give rather interesting information and may
be crucial to  test further developments of the theory. For example the
theory could  predict
 that in $d$-dimensions $\g=d-2$, i.e.
\be
\tau \propto \exp({A \over (T-T_G)^{(d-2)}}).
 \ee
The only possibility for a new testing this kind of predictions (apart from
the known three dimensional case) would be to do accurate numerical
simulations
in higher dimensions.

 The study of off equilibrium dynamics in glass related
models has started only recently. Very interesting results have been
obtained and I am confident that this approach will lead to a better
understanding of real glasses.

 \section{Acknowledgements} It is a pleasure for me to thank for many
useful discussions and the
very pleasant collaboration on these  problems L. Cugliandolo, J.
Kurchan, E. Marinari, F. Ritort
and M. Virasoro.

 \end{document}